\documentclass[numberedappendix,twocolappendix,apjl]{emulateapj}
\usepackage{ulem}
\usepackage{amsmath}
\usepackage{graphicx}
\usepackage{subfigure}
\usepackage{color}
\shorttitle{Cosmic Rays in the Galactic CMZ}
\shortauthors{Yoast-Hull, Gallagher, \& Zweibel}
\begin{document}
\title{The Cosmic Ray Population of the Galactic Central Molecular Zone}

\author{Tova M. Yoast-Hull$^{1,2}$, J. S. Gallagher III$^3$, and Ellen G. Zweibel$^{1,2,3}$}

\affil{$^1$Department of Physics, University of Wisconsin-Madison, WI, USA; email: {\tt yoasthull@wisc.edu}}
\affil{$^2$Center for Magnetic Self-Organization in Laboratory and Astrophysical Plasmas, University of Wisconsin-Madison, WI, USA}
\affil{$^3$Department of Astronomy, University of Wisconsin-Madison, WI, USA}

%Abstract

\begin{abstract}
The conditions in the Galactic Center are often compared with those in starburst systems, which contain higher supernova rates, stronger magnetic fields, more intense radiation fields, and larger amounts of dense molecular gas than in our own Galactic disk.  Interactions between such an augmented interstellar medium and cosmic rays result in brighter radio and $\gamma$-ray emission.  Here, we test how well the comparisons between the Galactic Center and starburst galaxies hold by applying a model for cosmic ray interactions to the Galactic Center to predict the resulting $\gamma$-ray emission.  The model only partially explains the observed $\gamma$-ray and radio emission.  The model for the $\gamma$-ray spectrum agrees with the data at TeV energies but not at GeV energies.  Additionally, as the fits of the model to the radio and $\gamma$-ray spectra require significant differences in the optimal wind speed and magnetic field strength, we find that the single-zone model alone cannot account for the observed emission from the Galactic Center.  Our model is improved by including a soft, additional cosmic-ray population.  We assess such a cosmic ray population and its potential sources and find that a cosmic-ray electron spectrum is energetically favored over a cosmic-ray proton spectrum.
\end{abstract}
\keywords{cosmic rays, galaxies: starburst, Galaxy: center, gamma rays: galaxies, radio continuum: galaxies}

%Introduction

\section{Introduction}

Despite its proximity and wealth of existing observations, the origin of the radio and $\gamma$-ray emission from the Galactic Center remains controversial.  Previous attempts at modeling the observed $\gamma$-ray spectrum have focused their efforts on unresolved point sources or dark matter \citep[e.g.][]{Abazajian11, Gordon13}.  Alternatively, \cite{YusefZadeh13} have attempted to explain both the radio and $\gamma$-ray emission as diffuse emission from cosmic-ray electrons.  Resolving these problems is especially critical given the similarities of the Galactic Center to starburst galaxies.  

Intense magnetic and radiation fields and concentrated molecular gas content in the Galactic Center are reminiscent of conditions in starburst systems.  Additionally, the slope of the local cosmic ray spectrum is steeper than its spectrum at its source due to the effects of diffusion and energy losses.  In denser environments the cosmic-ray electron spectral steepening is due only to energy losses and not energy-dependent diffusion.  This hypothesis has been tested in the past in starburst nuclei as both radio and $\gamma$-ray observations are available for a few such objects.  To test how well the parallels between the Galactic Center and starburst galaxies apply, we apply a model for cosmic ray interactions designed for starburst nuclei \citep[][hereafter YEGZ]{YoastHull13} to the Central Molecular Zone (CMZ) of the Galactic Center and include both cosmic-ray protons and electrons.

Concentrating on the recent $\gamma$-ray observations by \textit{Fermi} and HESS, we find that the starburst model explains the TeV energy emission quite well but underpredicts the emission at GeV energies.  We also explore what kind of cosmic ray population would be necessary to produce the observed GeV energy emission and what types of sources could account for these cosmic rays.  Finally, we model the radio spectrum and investigate the impact of the additional population of cosmic rays introduced to model the GeV energy $\gamma$-rays.

The next section provides a brief overview of how the population of energetic particles was computed and describes the observed properties of the Galactic Center.  Section 3 contains the results of our model for the Galactic Center and Section 4 provides a discussion of the implications and concluding remarks.

%GC Properties
\section{Theoretical Model \& Parameters}

%Theory
\subsection{YEGZ Model}

%Model Parameters Table
\begin{center}
\begin{deluxetable*}{lllc}
%
%\tabletypesize{\scriptsize}
%\rotate
\tablecaption{Input Model Parameters}
\tablewidth{0pt}
\tablehead{
\colhead{Physical Parameters} & \colhead{Model A} & \colhead{Model B} & \colhead{References}
}
\startdata
Distance & 8.0 kpc & 8.0 kpc & 1 \\
CMZ Radius & 200 pc & 250 pc & 2 \\
CMZ Disk Scale Height & 50 pc & 50 pc & 2 \\
Molecular Gas Mass & $3 \times 10^{7}$ $M_{\odot}$ & $5 \times 10^{7}$ $M_{\odot}$ & 3 \\
Ionized Gas Mass\tablenotemark{a} & $2.7 \times 10^{3}$ $M_{\odot}$ & $7.3 \times 10^{3}$ $M_{\odot}$ & \\
Average ISM Density\tablenotemark{b} & $\sim$80 cm$^{-3}$ & $\sim$85 cm$^{-3}$ & \\
FIR Luminosity & $4\times 10^{8}$ $L_{\odot}$ & $4\times 10^{8}$ $L_{\odot}$ & 3 \\
FIR Radiation Field Energy Density\tablenotemark{b} & 13 eV~cm$^{-3}$ & 8 eV~cm$^{-3}$ & \\
Dust Temperature & 21 K & 21 K & 3 \\
Stellar IR Luminosity & $2.5\times 10^{9}$ $L_{\odot}$ & $2.5\times 10^{9}$ $L_{\odot}$ & 3 \\
Stellar IR Radiation Field Energy Density\tablenotemark{b} & 82 eV~cm$^{-3}$ & 53 eV~cm$^{-3}$ & \\
Effective Stellar Temperature & 4400 K & 4400 K & 3 \\
Star-Formation Rate (SFR) & $0.01$ M$_{\odot}$ yr$^{-1}$ & $0.025$ M$_{\odot}$ yr$^{-1}$ & 4 \\
SN Explosion Rate ($\nu_{SN}$)\tablenotemark{b} & $10^{-4}$ yr$^{-1}$ & $2.75 \times 10^{-4}$ yr$^{-1}$ & \\
SN Explosion Energy\tablenotemark{c} & 10$^{51}$ ergs & 10$^{51}$ ergs & \\
SN Energy in Cosmic-Ray Protons\tablenotemark{c} & 10\% & 10\% & \\
Ratio of Primary Protons to Electrons ($N_{p}$/$N_{e}$) & 50 & 50 & \\
Slope of Primary Cosmic Ray Source Function & 2.3 & 2.3 & \\
\enddata
%
%\tablecomments{Comments here...}
%
\tablenotetext{a}{Scales with the star-formation rate and molecular gas mass}
\tablenotetext{b}{Derived from above parameters}
\tablenotetext{c}{Excludes neutrino energy}
\tablerefs{
(1)~\cite{Reid09}; (2)~\cite{Ferriere07}; (3)~\cite{Launhardt02}; (4)~\cite{Longmore13};
}
\end{deluxetable*}
\end{center}

\setcounter{footnote}{3}

Previously, we created a model for cosmic ray interactions in the CMZs of star-forming galaxies capable of calculating both the radio and $\gamma$-ray emission due to cosmic ray interactions (YEGZ).  Our treatment of cosmic ray interactions is similar to several other works \citep[e.g.][]{Torres04, Lacki10, Paglione12}.  As our single zone model is designed for starburst systems in which strong galactic winds are present and environmental conditions are extreme enough that the energy loss timescales are significantly less than diffusion timescales, we include only energy and advective losses.  Thus, following the approach in YEGZ, the spectrum for cosmic rays depends only on the injection spectrum and the lifetime.  We assume a power-law source function, $Q(E) \propto E^{-p}$, for cosmic rays such that
\begin{equation}
\int_{E_{\text{min}}}^{E_{\text{max}}} Q(E) E dE = \frac{\eta \nu_{\text{SN}} E_{51}}{V} ,
\end{equation}
where $\nu_{\text{SN}}$ is the volume integrated supernova rate, $V$ is the volume of the starburst region, $\eta$ is the fraction of the supernova energy transferred to cosmic rays, and $E_{51} = 1$ is 10$^{51}$ ergs, the typical energy from a supernova explosion.  The steady state cosmic-ray proton spectrum is given by
\begin{equation}
N(E) = \frac{(p-2)}{E_{\text{min}}^{-p+2}} ~ \frac{\eta \nu_{\text{SN}} E_{51}}{V} E^{-p} ~ \tau(E),
\end{equation}
where $E_{\text{min}}$ is the minimum cosmic ray energy, here taken to be $E_{\text{min}} = 0.1$ GeV, and $p$ is the spectral index.  The cosmic ray lifetime is determined by the combined radiative and collisional energy loss and advection (energy-independent) timescales
\begin{equation}\label{tau}
\tau(E)^{-1} \equiv \tau_{\text{adv}}^{-1} + \tau_{\text{loss}}^{-1} = \left( \frac{H}{v_{\text{adv}}}  \right)^{-1} + \left( - \frac{E}{dE/dt} \right)^{-1},
\end{equation}
where $H$ is the scale height of the starburst region and $v_{\text{adv}}$ is the speed of the particles in the wind from the starburst region \citep[c.f.][]{Torres04}.  Energy losses include ionization, the Coulomb effect, and pion production\footnote{Proton-proton collisions produce a variety of secondary mesons including pions and kaons.  However, in terms of secondary electron/positron and $\gamma$-ray production, pions dominate over other mesons produced in these interactions \citep{Dermer09}.  As such, we only consider pion production and decay in our calculations.} for cosmic-ray protons and ionization, bremsstrahlung, inverse Compton emission, and synchrotron emission for cosmic-ray electrons (see YEGZ for further details).

In our models, we assume a three-phase interstellar medium (ISM) composed of a diffuse, hot gas that fills the majority of the volume with clumps of warm, ionized gas and dense molecular gas.  Interactions between the cosmic-ray protons and the molecular gas result in the production of a variety of secondary pions which quickly decay.  Neutral pion decay into $\gamma$-rays is the main hadronic channel for $\gamma$-ray production.  Additionally, charged pions decay into secondary electrons and positrons.  Additionally, while pions can also be produced in proton-photon interactions in the presence of sufficiently intense radiation fields \citep{Schlick02}, the pion production rate from this process in the CMZ is negligible compared to that due to proton-proton interactions.

In addition to hadronic $\gamma$-ray processes, we include the leptonic production mechanisms of bremsstrahlung and inverse Compton.  We combine the primary cosmic-ray electron population with the secondary electron/positron population when calculating the emission from leptonic processes.  While $\gamma$-rays from bremsstrahlung and $\pi^{0}$ decay are relatively simple to model, $\gamma$-rays from inverse Compton are more difficult to calculate due to the knowledge required about the stellar and thermal infrared radiation fields with which cosmic-ray electrons and positrons interact.  We assume a modified, diluted blackbody radiation spectrum \citep[see][]{YoastHull14}, taking the radiation energy density and temperature from the observed far-infrared flux for the CMZ (see Table 1).

\begin{figure*}[t!]
%\epsscale{1.15}
 \subfigure[Model A, $U_{\text{rad, dust}} = 13$ eV~cm$^{-3}$]{
  \includegraphics[width=0.5\linewidth]{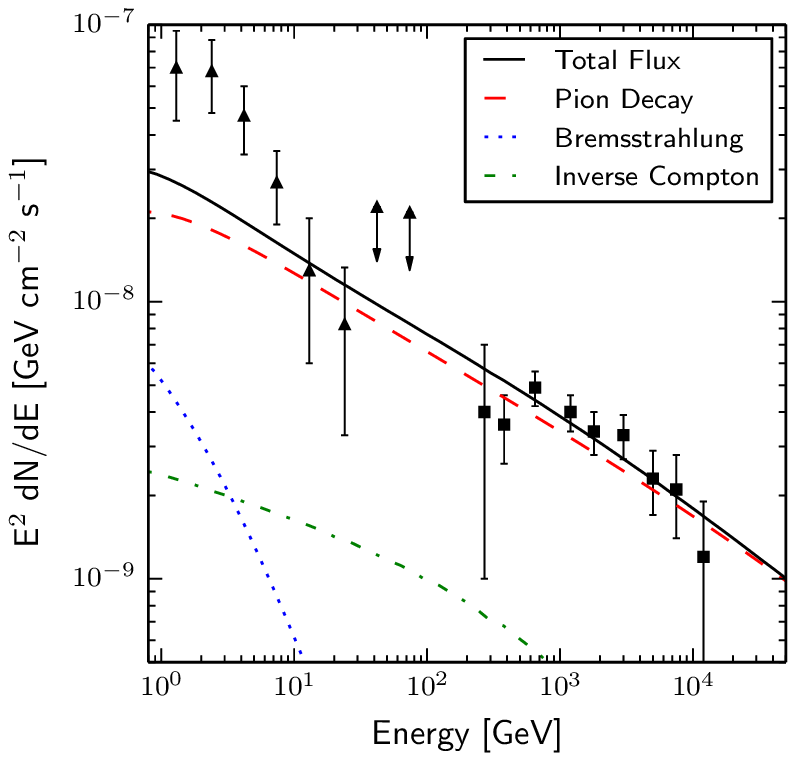}}
 \subfigure[Model B, $U_{\text{rad, dust}} = 8$ eV~cm$^{-3}$]{
  \includegraphics[width=0.5\linewidth]{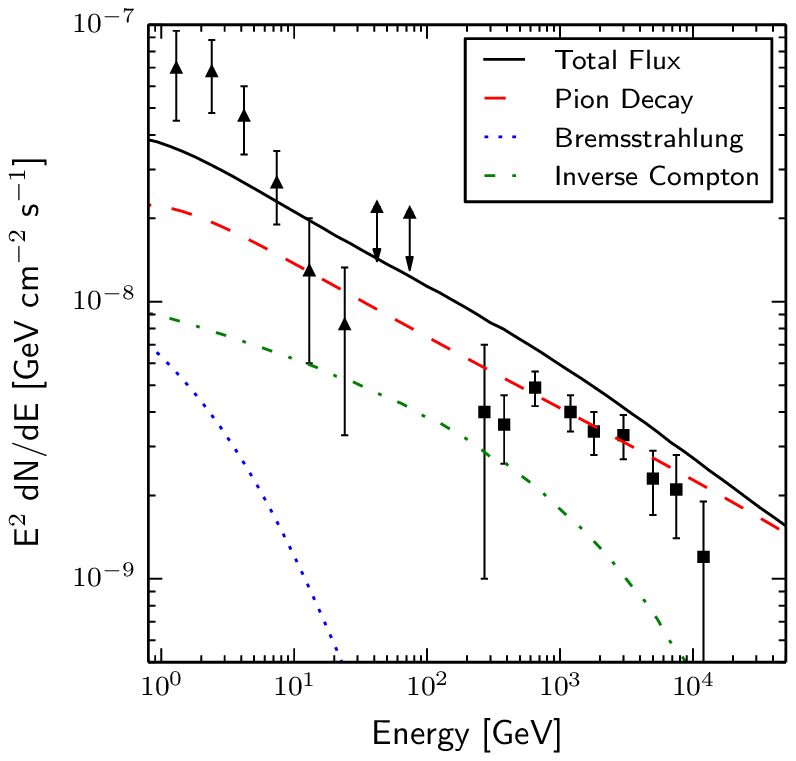}}
\caption{Best-fit $\gamma$-ray spectra for YEGZ Models A \& B.  Model parameters are set at $p = 2.3$, $B = 350$ $\mu$G with (\textit{left}) $v_{\text{adv}} = 700$ km~s$^{-1}$, $U_{\text{rad, dust}} = 13$ eV~cm$^{-3}$ and (\textit{right}) $v_{\text{adv}} = 2000$ km~s$^{-1}$, $U_{\text{rad, dust}} = 8$ eV~cm$^{-3}$.  The solid lines represent the total $\gamma$-ray flux; the dashed lines represent the contribution from neutral pion decay.  The dotted lines represent the contribution from bremsstrahlung and the dot-dashed lines represent the contribution from inverse Compton.  $\gamma$-ray data include: \cite{YusefZadeh13} (\textit{Fermi} - triangles), \cite{Aharonian06} (HESS - squares).  Data with downward arrows represent upper limits for both \textit{Fermi} and HESS data.}
\end{figure*}

Finally, in calculating the radio spectrum, we include synchrotron emission from the diffuse, hot gas from both primary cosmic-ray electrons and secondary cosmic-ray electrons and positrons from charged pion decay.  The effects of free-free emission and absorption from the warm, ionized gas are also included \citep[for details see][]{YoastHull14}.  While free-free emission is responsible for a significant portion of the radio emission at high frequencies, free-free absorption \citep[$\alpha_{\nu}^{ff} \propto \nu^{-2}$;][]{RL79} can flatten or completely turn down the radio spectrum at low frequencies \citep{Condon92}.

In the case of M82, the ionized gas acts as a foreground screen such that the radio emission from the starburst core is completely absorbed at low frequencies \citep{Adebahr13}.  Here, we leave the covering fraction ($f_{\text{abs}}$) for the ionized gas as a variable such that for a low covering fraction ($f_{\text{abs}} \sim 0.1 - 0.2$), only a small portion of the radio emission is absorbed by free electrons and the spectrum flattens \citep{YoastHull14} and for a large covering fraction ($f_{\text{abs}} \sim 1.0$), the radio emission turns over completely at low frequencies (YEGZ).

%GC Properties
\subsection{Properties of the Galactic Center}

In this paper, we consider only the region of the Galactic Center known as the central molecular zone (CMZ), spanning the inner $\sim$500 pc \citep{Launhardt02}.  This region is primarily characterized by its 180-pc radius molecular ring and a highly asymmetrical distribution of interstellar dust and gas \citep{Launhardt02, Jones13}.  Measurements of the total mass of the molecular gas range from $\sim10^{7}$ to $10^{8}$ M$_{\odot}$ with the majority of the molecular gas being contained in compact clouds occupying only a small percentage of the total volume \citep{Ferriere07}.  Here, we adopt a gas mass of $(3 - 5) \times 10^{7}$ M$_{\odot}$ similar to \cite{Ferriere07} (see Table 1) and assume the cosmic rays sample the mean density \citep{Boettcher13}.  Extensive infrared observations show that while the CMZ contains both cold (21 K) and warm (49 K) interstellar dust, the warm dust makes up only a small fraction of the dust by mass \citep{Launhardt02}.  Thus, when modeling the interstellar radiation field, we use a modified blackbody spectrum with a dust temperature of 21 K.  

\begin{figure*}[t!]
%\epsscale{1.15}
 \subfigure[Model A]{
  \includegraphics[width=0.5\linewidth]{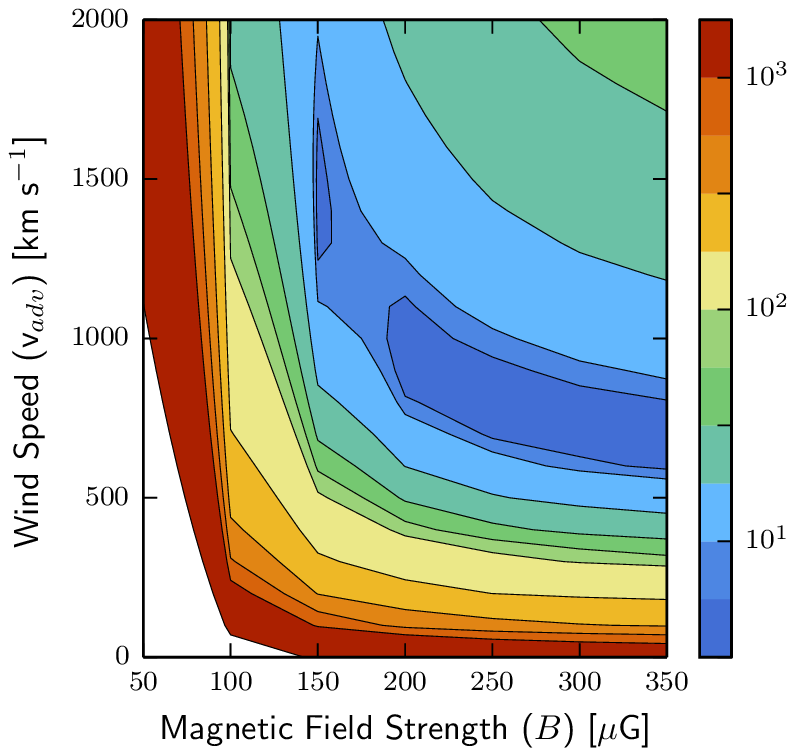}}
 \subfigure[Model B]{
  \includegraphics[width=0.5\linewidth]{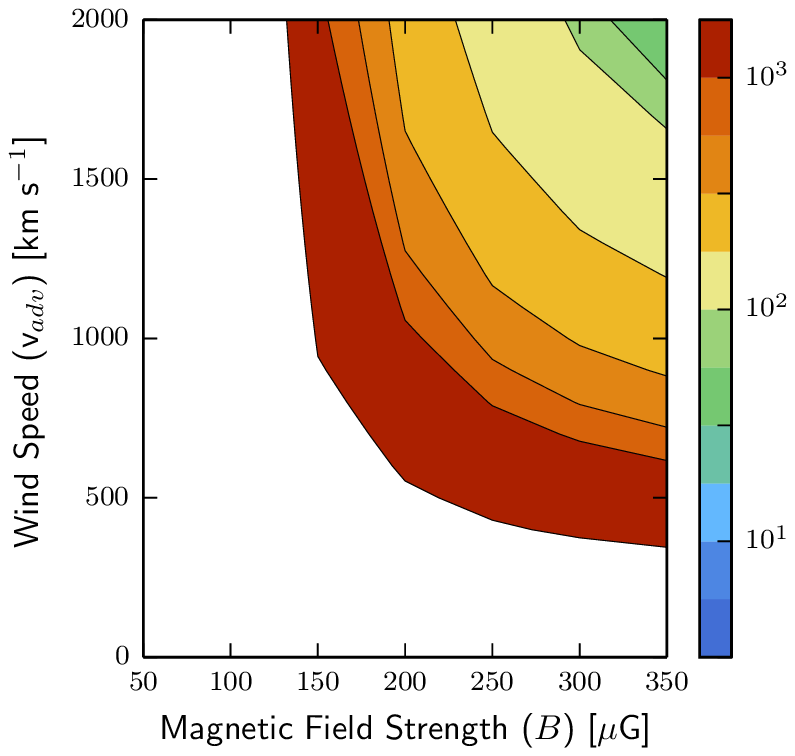}}
 \subfigure[Model A, Soft Electron Spectrum]{
  \includegraphics[width=0.5\linewidth]{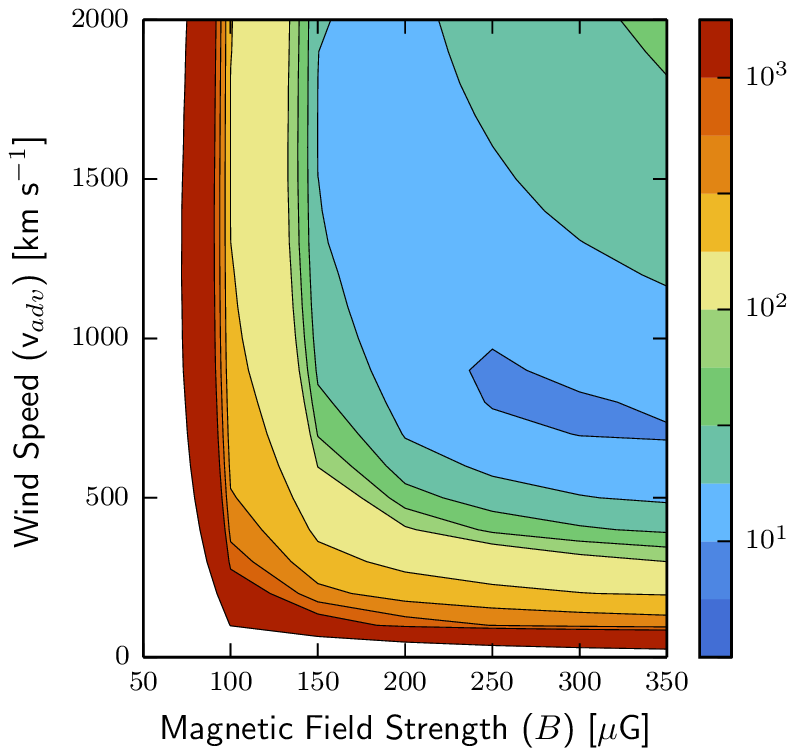}}
 \subfigure[Model A, Soft Proton Spectrum]{
  \includegraphics[width=0.5\linewidth]{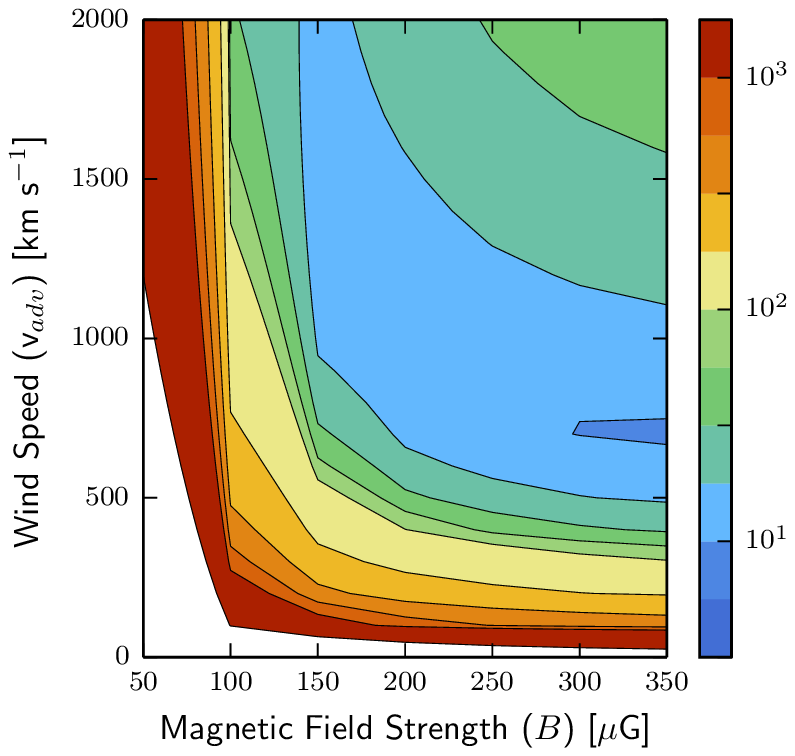}}
\caption{Contour plots showing $\chi^{2}$ variations for fits to the $\gamma$-ray spectrum for the range of magnetic field strength ($B$) and advection (wind) speed ($v_{adv}$) with parameters for Models A (top left \& bottom) and B (top right) listed in Table 1. For the top plots, the total number of degrees of freedom is 9 and it is 15 for the bottom plots.  The top plots show the shape of parameter space for a single cosmic ray spectrum while the bottom plots include an extra soft spectrum of electrons with $p = 2.7$ (left) and protons with $p = 3.1$ (right).}
\end{figure*}

Other characteristic properties of the Galactic Center include its supermassive black hole \citep[M$_{\text{SMBH}} = 4.4 \times 10^{6}$ M$_{\odot}$;][]{Genzel10} which is currently inactive, as evidence for a radiating accretion disk is lacking \citep{Mezger96}.  However, while structures such as the Fermi bubbles may be indicators of past activity \citep{Guo12, Yang12}, the majority of the emission seen from the inner $\sim$30 pc is due to the dense complex of star-formation.  Based on measurements of the far-infrared luminosity and ionization rates for the CMZ, we adopt star-formation rates in the range of $0.01 - 0.025$ M$_{\odot}$ yr$^{-1}$ \citep{Longmore13}.

Radio observations of the CMZ also reveal a highly asymmetrical structure with a mixture of thermal and non-thermal sources \citep{Law08}.  Several non-thermal filamentary structures are found throughout the region and likely arise from nearby star clusters \citep{Law08}.  Additionally, these filaments are tracers of the organized magnetic field perpendicular to the Galactic disk \citep{YusefZadeh13}.  Estimates of the magnetic field strength range from $\sim$10 $\mu$G \citep{LaRosa05, YusefZadeh13} to as high as $\sim$1 mG \citep{YusefZadeh84, Morris96}.  We consider magnetic field strengths of a few hundred $\mu$G as in nearby starburst environments \citep[e.g.][]{Crocker11, Thompson07}.

$\gamma$-ray observations at both GeV and TeV energies exist for the CMZ.  Diffuse $\gamma$-ray observations of the CMZ at TeV energies have been shown to be coincident with a large complex of molecular clouds by the HESS collaboration \citep{Aharonian06}.  It is likely that these TeV $\gamma$-rays were produced in interactions between cosmic rays and the interstellar gas; however, it is unclear exactly what is responsible for the diffuse GeV energy emission \citep[see][and references therein]{Gordon13}.

%Best-Fits Table
\begin{center}
\begin{deluxetable*}{lccccccccccc}
%
%\tabletypesize{\scriptsize}
%\rotate
\tablecaption{Best-Fit Parameters}
\tablewidth{0pt}
%
%\tablehead{
%\colhead{Data Set} & \colhead{Extra CRs} & \colhead{Model} & \colhead{$\chi^{2}$} & \colhead{$p$} & \colhead{$B$ ($\mu$G)} & \colhead{$v_{\text{adv}}$ (km~s$^{-1}$)} & \colhead{$n_{\text{ion}}$ (cm$^{-3}$)} & \colhead{$f_{\text{abs}}$} & \colhead{\# Data Points} & \colhead{Energy in CRs (ergs)} 
%}
%
\tablehead{
\colhead{Data} \vspace{-0.15cm} & \colhead{Extra} & & & & \colhead{$B$} & \colhead{$v_{\text{adv}}$} & \colhead{$n_{\text{ion}}$} & & \colhead{\# of} & & \colhead{$E_{CR} \times \nu / \nu_{SN}$}\\ \vspace{-0.15cm}
& & \colhead{Model} & \colhead{$\chi^{2}$} & \colhead{$p$} & & & & \colhead{$f_{\text{abs}}$} & & \colhead{d.o.f.} & \\
\colhead{Set} & \colhead{Component} & \colhead{} & \colhead{} & \colhead{} & \colhead{($\mu$G)} & \colhead{(km~s$^{-1}$)} & \colhead{(cm$^{-3}$)} & \colhead{} & \colhead{Data Points} & \colhead{} & \colhead{($10^{50}$ ergs)}
}
\startdata
TeV $\gamma$-Rays & -- & A & 5.3 & 2.3 & 350 & 700 & -- & -- & 9 & 7 & 1.02\\ 
TeV $\gamma$-Rays & -- & B & 52.9 & 2.3 & 350 & 2000 & -- & -- & 9 & 7 & 1.02\\ \\
All $\gamma$-Rays & -- & A & 19.2 & 2.3 & 300 & 700 & -- & -- & 15 & 13 & 1.02\\ 
All $\gamma$-Rays & -- & B & 64.4 & 2.3 & 350 & 2000 & -- & -- & 15 & 13 & 1.02\\ \\
All $\gamma$-Rays & Electrons & A & 9.2 & 2.7 & 250 & 900 & -- & -- & 15 & 13 & 1.84\\
All $\gamma$-Rays & Electrons & B & 61.5 & 2.9 & 350 & 2000 & -- & -- & 15 & 13 & 3.96\\ \\
All $\gamma$-Rays & Protons & A & 9.2 & 3.1 & 350 & 700 & -- & -- & 15 & 13 & 174\\
All $\gamma$-Rays & Protons & B & 61.0 & 3.3 & 350 & 2000 & -- & -- & 15 & 13 & 254\\ \\
Radio & -- & A & 169 & 2.3 & 200 & 100 & 100 & 1.0 & 4 & 0 & 1.02\\ 
Radio & -- & B & 70.7 & 2.3 & 250 & 500 & 100 & 1.0 & 4 & 0 & 1.02\\ \\ 
Radio & Electrons & A & 73.4 & 2.7 & 100 & 300 & 75 & 1.0 & 4 & 0 & 1.38\\ 
Radio & Electrons & B & 60.0 & 2.3 & 100 & 300 & 75 & 1.0 & 4 & 0 & 1.05\\ \\
Radio \& $\gamma$-rays\tablenotemark{a} & -- & A & 342 & 2.3 & 350 & 300 & 100 & 1.0 & 19 & 15 & 1.02\\ 
Radio \& $\gamma$-rays\tablenotemark{a} & -- & B & 396 & 2.3 & 350 & 1400 & 100 & 1.0 & 19 & 15 & 1.02\\ \\
Radio \& $\gamma$-rays\tablenotemark{a} & Electrons & A & 572 & 2.7 & 100 & 700 & 75 & 1.0 & 19 & 15 & 1.48\\ 
Radio \& $\gamma$-rays\tablenotemark{a} & Electrons & B & 2980 & 2.7 & 150 & 1900 & 50 & 1.0 & 19 & 15 & 1.40\\ \\
\enddata
\tablenotetext{a}{While we list the combined solutions for the $\gamma$-ray and radio spectra, the results are heavily weighted by the radio spectrum and the corresponding $\gamma$-ray $\chi^{2}$ values are $\gtrsim$200.  Thus, there is no optimal solution that agrees well with both data sets.}
\tablecomments{The final column of the table is the energy input into cosmic rays (per supernovae).  In the original starburst model, $10^{50}$ erg is put into cosmic-ray protons and $2 \times 10^{48}$ erg goes into cosmic-ray electrons.  Thus, for the original models, the last column reads 1.02.  For the rows with an additional soft spectrum, the additional energy is the amount required to match the soft spectrum with the observed GeV energy data.}
\end{deluxetable*}
\end{center}
%

%Results
\section{Results}

%Gamma Rays
\subsection{TeV $\gamma$-Ray Spectrum}

Previously, we used $\chi^{2}$ tests to optimize the input values for magnetic field strength ($B$), wind speed ($v_{\text{adv}}$), ionized gas density ($n_{\text{ion}}$), and absorption fraction ($f_{\text{abs}}$).  While magnetic field strength and wind speed affect both the $\gamma$-ray and radio spectra, ionized gas density and absorption fraction primarily affect the radio spectrum and have little to no effect on the predicted $\gamma$-ray spectrum.  We also explore how uncertainties in observational properties of the CMZ (supernova rate, CMZ radius, and molecular gas mass) affect minimization.  We test two scenarios for a minimum (Model A) and a maximum (Model B) set of parameters (see Table 1).

For Model A, we find that our model accurately predicts the TeV $\gamma$-ray emission but underestimates the majority of the observed GeV energy data (see Figure 1).  At TeV energies, most of the modeled emission is due to $\pi^{0}$ decay with a rather small contribution from inverse Compton while bremsstrahlung makes a significant contribution at GeV energies only.  A significant wind with speeds of $\sim$600 -- 900 km~s$^{-1}$ is required to avoid overestimating the TeV spectrum.  

Models within one sigma of the best-fit model have magnetic field strengths ranging from 150 to 350 $\mu$G.  While one might expect the $\gamma$-ray spectrum to be independent of magnetic field strength, inverse Compton and synchrotron are competitive processes.  As magnetic field strength increases, electrons are emitting more synchrotron radiation, leaving less energy available for inverse Compton.  Thus, to achieve the same goodness of fit, the wind speed must decrease with increasing magnetic field strength such that the electron spectrum remains the same (see Figure 2).  This is consistent with our previous results for the $\gamma$-rays in other galaxies and is opposite the relationship found for the radio spectrum \citep{YoastHull14}.  Additionally, while the observational values assumed for Model A easily lead to a best-fit model, the $\chi^{2}$ results for Model B never minimize.  Thus, Model B is not a suitable model (see Figure 2 \& Table 2).  

\begin{figure*}[t!]
%\epsscale{1.15}
 \subfigure[Model A, Soft Electron Spectrum]{
  \includegraphics[width=0.5\linewidth]{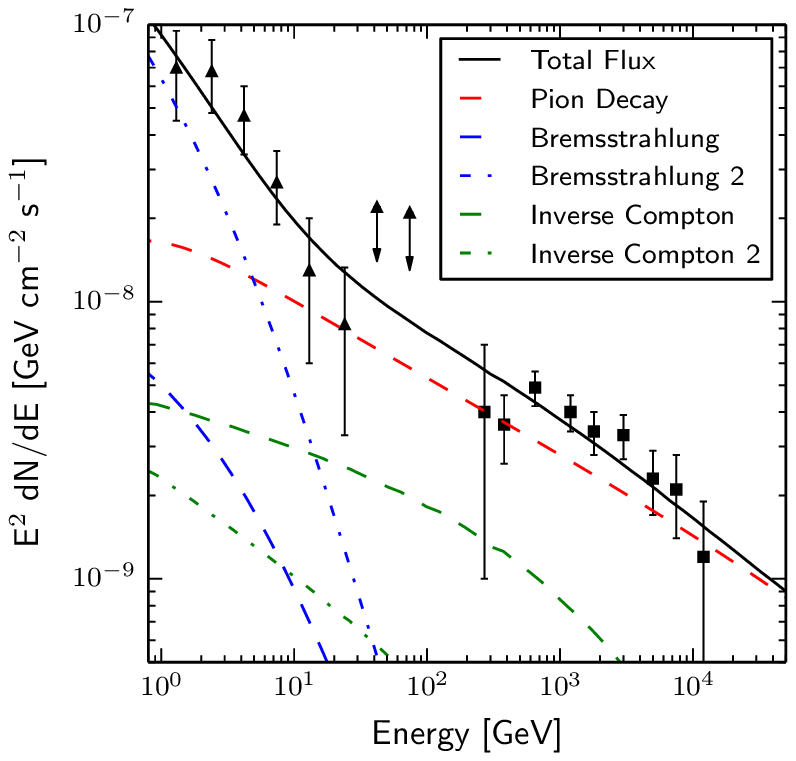}}
 \subfigure[Model A, Soft Proton Spectrum]{
  \includegraphics[width=0.5\linewidth]{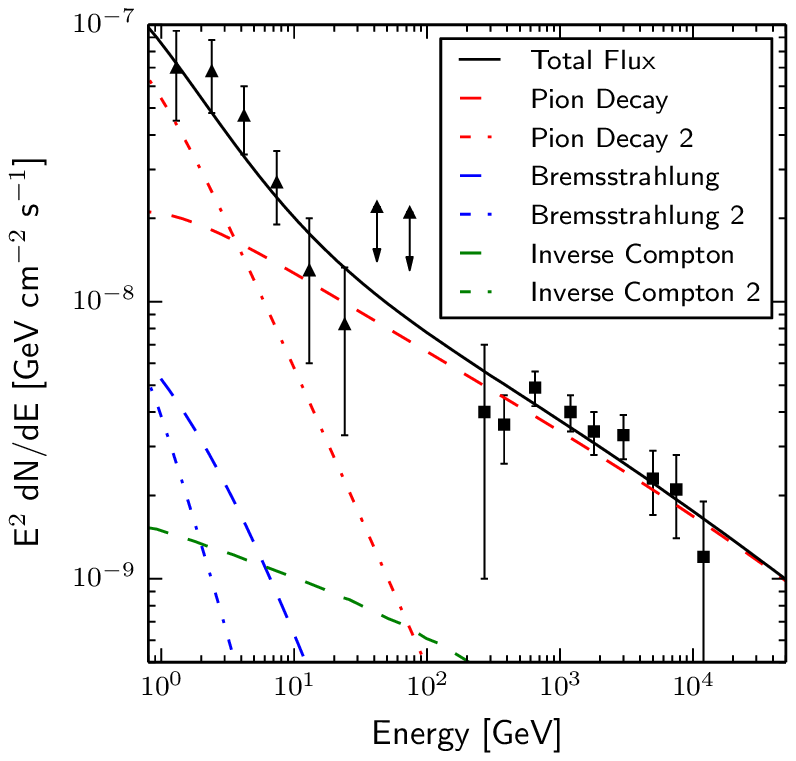}}
\caption{Best-fit $\gamma$-ray spectra including additional soft component.  Model parameters are set at (\textit{left}) $p = 2.7$ for an additional soft electron spectrum with $B$ =  and (\textit{right}) $p = 3.1$ for an additional soft proton spectrum.  The soft proton spectrum includes inverse Compton and bremsstrahlung from secondary electrons/positrons (as both neutral and charged pions are produced in proton-proton collisions) and the soft electron spectrum includes emission from both bremsstrahlung and inverse Compton.  Dashed lines show emission from the original starburst model while dot-dashed lines, labeled as ``2,'' show emission from the additional soft component.  $\gamma$-ray data include: \cite{YusefZadeh13} (\textit{Fermi} - triangles), \cite{Aharonian06} (HESS - squares).  Data with downward arrows represent upper limits for both \textit{Fermi} and HESS data.}
\end{figure*}

As the average densities for the two models are essentially the same, the difference in results must be attributed to the choice in supernova rate or acceleration efficiency, as only the product of the two can be constrained.  The changes caused by the choice in supernova rate are seen, in particular, in the contribution of bremsstrahlung and inverse Compton to the total $\gamma$-ray flux in Figure 1.  The $\gamma$-ray flux from leptons increases by a factor of $\sim$3 from Model A to Model B while the flux from hadronic interactions is essentially the same.  To ensure that the flux from neutral pion decay continues to agree with the TeV energy data, the best-fit for Model B requires an increase in wind speed to 2000 km~s$^{-1}$ from 700 km~s$^{-1}$ in Model A.  However, because the lifetimes of the cosmic-ray electrons are much shorter than that of the cosmic-ray protons, due to larger energy losses, the electron spectrum remains virtually unchanged by the increase in wind speed.  Thus, the increase in the flux from inverse Compton is not because of the difference in radiation field energy density or lifetime but the higher supernova rate.

In testing Models A \& B, we vary supernova rate, CMZ radius, and molecular gas mass.  While the far-infrared luminosity is essentially the same for both models, the change in volume means a change in radiation field energy density.  Observations of dust in the CMZ give a far-infrared luminosity of $L = 4 \times 10^{8}$ $L_{\odot}$ with a corresponding dust temperature of $\sim$21 K \citep{Launhardt02}.  Based on this luminosity and depending on assumptions about volume, the radiation field energy density ranges from $\sim$8 -- 13 eV~cm$^{-3}$ between Models A \& B.  In a typical starburst, the radiation field due to far-infrared emission from dust would be the dominant field.  However, in the Galactic Center, the stellar radiation field is actually larger than that from dust.  Observations show that infrared emission due to stars has a luminosity of $L = 2.5 \times 10^{9}$ $L_{\odot}$ with a temperature of $\sim$4400 K \citep{Launhardt02}.  This corresponds to a radiation field energy density of $\sim$53 -- 82 eV~cm$^{-3}$.

Other models for the Galactic Center include \cite{Lacki13} and \cite{Crocker11}, each with different approaches to the radiation field.  \cite{Lacki13} use the far-infrared luminosity in a smaller volume ($R \sim 100$ pc) with a background radiation field which gives a total energy density of $\sim$60 eV~cm$^{-3}$.  \cite{Crocker11} assume an energy density of 90 eV~cm$^{-3}$ by taking the infrared luminosity in both dust and stars \citep[$L_{\text{TIR}} = 3.6 \times 10^{9}$ L$_{\odot}$;][]{Launhardt02}.  Though we treat the stellar and dust components separately in our calculation, it is unclear if \cite{Crocker11} do the same.  This makes little difference at lower energies for the energy losses.  However, the energy at which Klein-Nishina losses dominate such that inverse Compton is no longer a competitive energy loss mechanism depends on the radiation field temperature and thus different for each component.  The $\gamma$-ray emissivity for inverse Compton scattering also depends on temperature, not just energy density.  Thus, while interactions with infrared emission from dust result in a small contribution to the total $\gamma$-ray emissivity, scattering via starlight is completely negligible as the resulting inverse Compton emission is several orders of magnitude below that from the thermal infrared emission from dust.

\begin{figure*}[t!]
%\epsscale{1.15}
 \subfigure[Model B, Best-Fit Original Radio Spectrum]{
  \includegraphics[width=0.5\linewidth]{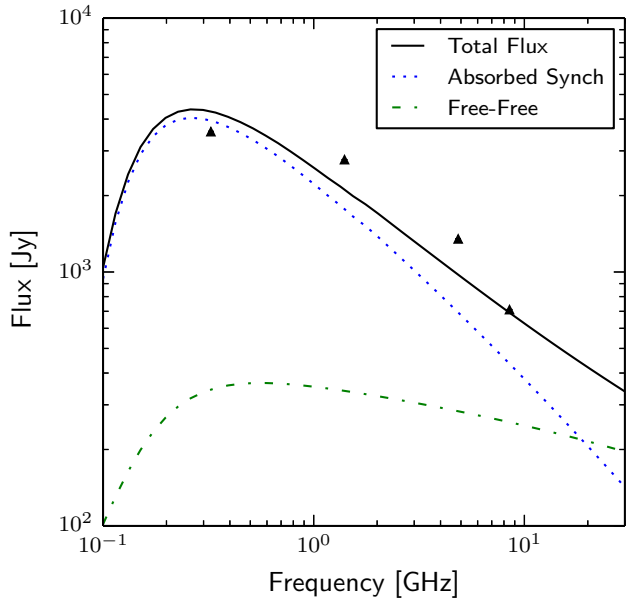}}
 \subfigure[Model A, Original Radio Spectrum ($\gamma$-Ray Parameters)]{
  \includegraphics[width=0.5\linewidth]{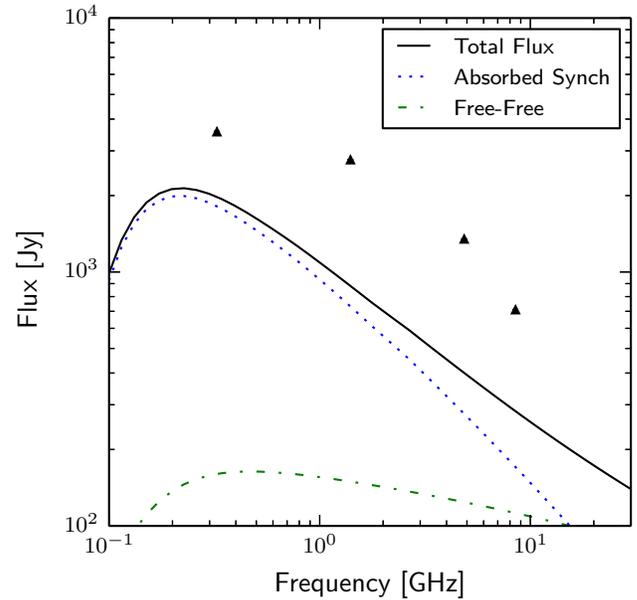}}
 \subfigure[Model B, Best-Fit for Radio Spectrum with Extra Comp.]{
  \includegraphics[width=0.5\linewidth]{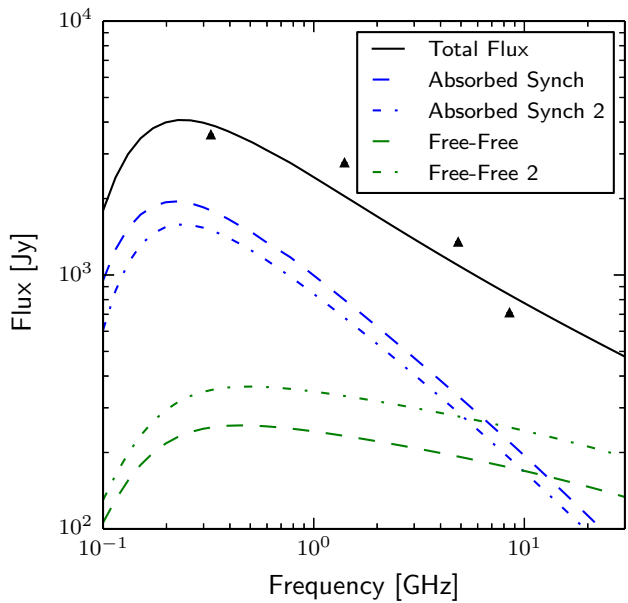}}
 \subfigure[Model A, Radio Spectrum with Extra Comp.  ($\gamma$-Ray Parameters)]{
  \includegraphics[width=0.5\linewidth]{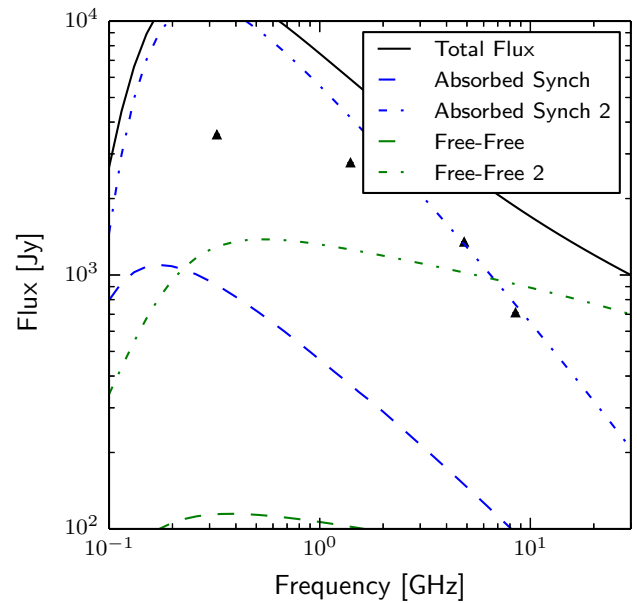}}
\caption{On the left are the best-fit radio spectra with (bottom) and without (top) an additional soft electron component.  Models on the right are the radio spectra for the best-fit $\gamma$-ray parameters with (bottom) and without (top) an additional soft electron component.  Parameters are list in Table 2 in rows 1 (top right), 5 (bottom right), 10 (top left), and 12 (bottom left).  For models at the bottom, dashed lines show emission from the original starburst model while dot-dashed lines show emission from the additional soft component.  Radio data come from \cite{YusefZadeh13}.}
\end{figure*}
%
%

%Gamma Rays
\subsection{GeV $\gamma$-Ray Spectrum}

The most notable feature of Figure 1 is that the starburst model matches up well with the $\gamma$-ray spectrum observed at TeV energies but is insufficient at GeV energies.  Not only does the model seriously underestimate the observed GeV emission, the spectral index of the model is significantly flatter than observed.  The ``excess'' of the observed Galactic Center $\gamma$-ray emission at GeV energies above typical background models is well-documented \citep[e.g.][]{Abazajian11, Crocker11, YusefZadeh13, Gordon13}.  Though there are many different explanations for this excess, ranging from an unresolved population of millisecond pulsars to dark matter annihilation, we focus only on the characteristics of a cosmic ray population necessary to account for the excess emission.  Additionally, we suggest what type of accelerator would be necessary to produce such a population.

We add a simple extra population of cosmic rays to our existing model for the $\gamma$-ray emission at TeV energies.  For the source function, we require the energy put into cosmic rays for the extra population be large enough that the model matches the observed emission at 1.3 GeV while leaving the average density ($n_{\text{ISM}}$), radiation field energy density ($U_{\text{rad}}$), and acceleration efficiency ($\eta$) unchanged.  Again, we use $\chi^{2}$ tests to determine the spectral index ($p$) required for a soft spectrum to match the GeV $\gamma$-ray emission while still optimizing magnetic field strength and wind speed for all cosmic rays.  As the model changes with wind speed and magnetic field strength, the required energy input also changes.  We test both a soft electron and a soft proton spectrum (including secondary electrons/positrons produced in proton-proton interactions but no additional primary electrons, see Figure 3).  Complete results for the $\chi^{2}$ tests can be found in Table 2.

\begin{figure*}[t!]
%\epsscale{1.15}
 \subfigure[Model B, Original Starburst Model]{
  \includegraphics[width=0.5\linewidth]{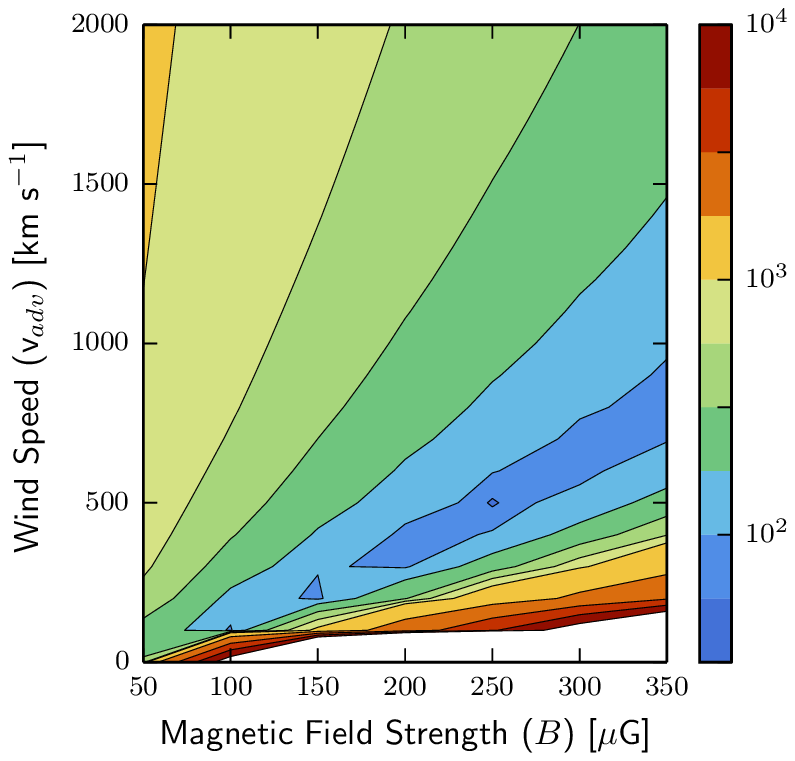}}
 \subfigure[Model B, Additional Soft Electron Component]{
  \includegraphics[width=0.5\linewidth]{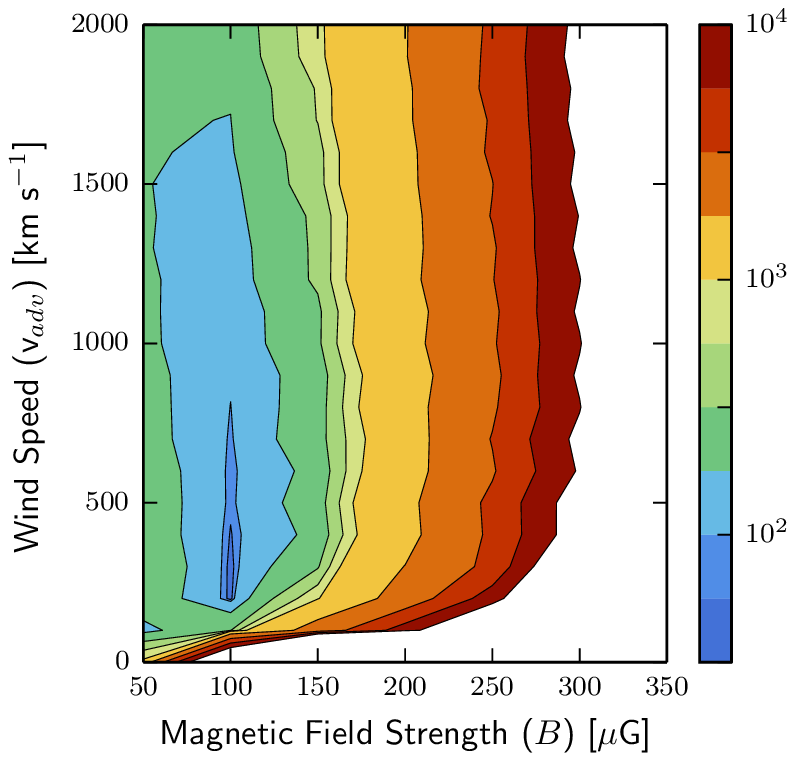}}
\caption{Contour plots showing $\chi^{2}$ values for the radio spectrum for varying magnetic field strength ($B$) and advection (wind) speed ($v_{adv}$) assuming parameters for Model B as listed in Table 1. For fits to the CMZ radio spectrum, the left plot shows the variation in $\chi^{2}$ for a single cosmic ray spectrum while the right plot includes an extra soft spectrum of electrons with $p = 2.3$.  For both plots, the total number of degrees of freedom is 0, as there are as many data points as free parameters.}
\end{figure*}

Results for the optimization of the magnetic field strength and wind speed are similar to previous results for only TeV emission (see Figure 2 \& Table 2), with a similar range of value for models within one sigma of the best-fit, though there is a more limited range for magnetic field strength.  The best-fit models with the extra cosmic ray populations have spectral indices of $p = 2.7$ for a soft electron spectrum and $p = 3.1$ for a soft proton spectrum.  For a standard supernova accelerator, the typical energy input into cosmic rays is $10^{50}$ erg for protons (assuming a 10\% acceleration efficiency) and $2 \times 10^{48}$ erg for electrons (assuming an electron-to-proton ratio of 0.02) \citep{Blandford78}.  The energy input per event is effectively the amplitude of the source function by which we scale our additional population, see Equation (1).  In fitting, we require the soft cosmic-ray spectrum to agree with the observed GeV data.  To accomplish this, we find that the soft electron spectrum needs 40 -- 150 times more energy per event than a standard model.  Thus, the total energy input into cosmic rays per event ranges from $1.84 \times 10^{50}$ erg to $3.96 \times 10^{50}$ erg.  The soft proton spectrum needs $\sim$200 times more energy per event, which is equivalent to a total energy input per event of $\sim 2 \times 10^{52}$ erg.  As such, the soft proton spectrum is excluded if supernovae are the source.

While cosmic-ray protons are accelerated in larger numbers than cosmic-ray electrons, the energy necessary for a soft proton component is significantly more than for a soft electron component due to differences in particle rigidity and radiative efficiency.  Additionally, if the GeV excess is a true bump, as suggested by \cite{Daylan14}, then $\gamma$-ray emission by neutral pion decay could be a more appropriate match to the spectrum shape.  However, we rule out a soft proton spectrum due to energy requirements and focus our efforts on an extra cosmic-ray electron population, as suggested by others \citep[e.g.][]{YusefZadeh13}.

%Radio
\subsection{Radio Spectrum}

In addition to modeling the $\gamma$-ray spectrum for the Galactic Center, we also model the radio spectrum as it is the main diagnostic we have for the cosmic-ray electron population.  However, we have as many data points as we have variables for the radio spectrum.  Thus, while the results for the radio spectrum will only give us a general idea of the optimal parameter space and not definitively constrain magnetic field strength or wind speed, they will provide a vital check on our additional soft electron spectrum.

We model the radio spectrum both with and without the additional soft electron population but discard the soft proton population.  Along with varying magnetic field strength ($B$) and wind speed ($v_{\text{adv}}$), we also vary ionized gas density ($n_{\text{ion}}$), from 25 to 100 cm$^{-3}$, and absorption fraction ($f_{\text{abs}}$), from 0.1 to 1.0, as they directly affect the amount of free-free absorption and emission.

The limitations of our single-zone model can most easily be seen in the radio spectrum.  Though the general level of flux observed can be achieved with our models, the specific shape of the spectrum is not, as seen in Figure 4.  Results for the original starburst model show that, in contrast to the $\gamma$-ray spectrum, the radio spectrum favors Model B and a lower magnetic field strength of $\sim$250 $\mu$G.  Again, we find a degeneracy between magnetic field strength and wind speed as seen in Figure 5.  To achieve a fit of similar goodness, wind speed must be increased as magnetic field increases so as to not overestimate the radio spectrum.  

With the addition of an extra soft electron spectrum, the goodness of the fit for Model A improves significantly while Model B remains largely the same (see Table 2).  However, even with the introduction of the soft component, Model B is still preferred to Model A.  Comparing the parameters for the best-fits for the radio spectrum and the $\gamma$-ray spectrum, we find very different wind speeds and magnetic field strengths for each.  As we have as many parameters as data points for the radio spectrum, the results are primarily used as a check on the $\gamma$-ray results.  In the case of the original starburst model, the minimized parameters for the $\gamma$-ray spectrum result in an underestimation of the radio spectrum (see Figure 4).  Conversely, for the extra soft component, not only does the preferred spectral index for the $\gamma$-rays not match with the radio data, but the radio spectrum is overestimated for the $\gamma$-ray parameters.  

%Conclusions
\section{Discussion \& Conclusions}

Applying our YEGZ model for starburst galaxies to the CMZ of the Galactic Center, we find that the model agrees well with the observed TeV energy $\gamma$-ray emission but that the model underpredicts the GeV energy emission.  While the model also agrees with the radio spectrum moderately well, our $\chi^{2}$ minimization shows that the $\gamma$-ray and radio spectra favor two different supernova rates.  While we can only constrain the product of supernova rate and acceleration efficiency, the ratio of primary protons to electrons is set by rigidity.  Thus, as the $\gamma$-rays at TeV energies are mainly produced in neutral pion decay and the radio spectrum is mostly from primary electrons with a relatively small contribution from secondaries, it is unlikely that the difference in requirements for the radio and $\gamma$-ray spectra are due to differences in acceleration efficiency.

Additionally, the results for the radio spectrum show that our best-fit models have a magnetic field strength of $\sim$100 -- 250 $\mu$G.  This is consistent with estimates by \cite{Crocker10}.  Of course, as we have as many data points as constraints, this is only a rough estimate of the field.  However, $\chi^{2}$ results show that the fits worsen as magnetic field strength decreases, and so it is unlikely that the magnetic field strength is on the order of $\sim$10 $\mu$G as suggested by \cite[][]{YusefZadeh13}.  Additional data and a more sophisticated model, however, are necessary to truly determine the magnetic field strength in the Galactic Center.

Finally, inclusion of an additional soft cosmic ray spectrum permits a fit to the GeV energy data.  The extra energy required rules out a soft proton spectrum in favor of a soft electron spectrum.  General possibilities for the source of any population of cosmic-ray electrons include SNRs, pulsar wind nebulae (PWNe), and the massive central black hole.  In regards to acceleration of a soft electron spectrum by supernovae, the amount of additional energy needed is achievable within the uncertainty for the supernova rate.  However, inclusion of this extra component results in a gross overestimation of the radio spectrum for parameters corresponding to the best-fit $\gamma$-ray models.  As such, the best-fit models for the radio spectrum are in very different areas of parameter space for both supernova rate and wind speed from the best-fits to the $\gamma$-ray spectrum (see Figures 2c and 5b).  This discrepancy between our YEGZ single-zone models suggests that the excess in the observed GeV data is unlikely to be diffuse in nature.

%Acknowledgements
\acknowledgements

This work was supported in part by NSF AST-0907837, NSF PHY-0821899 (to the Center for Magnetic Self-Organization in Laboratory and Astrophysical Plasmas), and NSF PHY-0969061 (to the IceCube Collaboration).  We thank the referee for their comments and help in improving our manuscript.  We acknowledge discussions with Roland Crocker and thank Francis Halzen for his help and support.  We also thank the organizers of IAU Symposium 303 on the Galactic Center for allowing this topic to be discussed.

%Bibliography

%
\end{document}